\documentclass[12pt]{article} 

\usepackage{amsfonts}
\usepackage{epsfig}
\usepackage{latexsym}
\usepackage{graphicx}
\def\bequ{\begin{equation}}
\def\eequ{\end{equation}}
\def\barr{\begin{array}}
\def\earr{\end{array}}

\def\ben{\begin{equation}}
\def\een{\end{equation}}
\def\bena{\begin{eqnarray}}
\def\eena{\end{eqnarray}}

\def\nn{\nonumber}
\newcommand{\sect}[1]{\setcounter{equation}{0}\section{#1}}

\def\thefootnote{\fnsymbol{footnote}}
\renewcommand{\thefootnote}{\alph{footnote}}

\newcommand{\beqn}{\begin{equation}}
\newcommand{\eeqn}{\end{equation}}
\newcommand{\beqarr}{\begin{eqnarray}}
\newcommand{\eeqarr}{\end{eqnarray}}

\newcommand{\matc}{\begin{array}{c}}
\newcommand{\matcc}{\begin{array}{cc}}
\newcommand{\matccc}{\begin{array}{ccc}}
\newcommand{\matcccc}{\begin{array}{cccc}}
\newcommand{\emat}{\end{array}}

\begin{document}

\begin{titlepage}

October 2003         \hfill
\begin{center}

\vskip .90in
\renewcommand{\thefootnote}{\fnsymbol{footnote}}
{\large \bf Over-Rotating Supertube Backgrounds}
\vskip .30in

Daniel Brace\footnote{email address: brace@physics.technion.ac.il}
\vskip .30in
{\it Department of Physics, Technion},
\\ {{\it Israel Institute of Technology}},
\\ {{\it Haifa 32000, Israel}}

\end{center}
\vskip 2.0in

\begin{abstract}
We study classical supertube probes on supergravity backgrounds which are sourced
by over-rotating supertubes, and which therefore contain closed timelike curves. 
We show that the BPS probes are stable despite the appearance
of negative kinetic terms in the probe action. By studying the radial oscillations of these probes, we show
that closed geodesics exist on these backgrounds.
\end{abstract}

\end{titlepage}

\newpage
\renewcommand{\thepage}{\arabic{page}}

\setcounter{page}{1}
\setcounter{footnote}{0}



\sect{Introduction}

In this paper, we study a class of backgrounds sourced by supertubes  \cite{MT}. These backgrounds were
constructed in \cite{Emparan:2001ux}, where it was also shown that if the angular momentum of the source 
violates a certain bound, the background will contain closed timelike curves. In addition, a 
supertube 
probe computation on these over-rotating backgrounds revealed that the kinetic terms of the probe action are
negative in certain regions and for a certain range of probe charges. This was interpreted as an instability
of the BPS supertube\footnote{ \lq BPS supertube' may seem redundant, but in this paper we will refer to a cylindrical
$D2$ brane in a state of radial oscillation as a supertube, despite the fact that this state is not supersymmetric.} 
probe, and was taken as evidence that the background is unphysical. 

In apparently unrelated work \cite{Gauntlett:2002nw} it was shown that low energy string
theory admits supersymmetric solutions of the G\"odel type \cite{Godel:ga}.
These homogeneous spaces also have closed timelike curves, but it was then pointed out that
they are not present in dimensional
upliftings of certain dual versions of these G\"odel spaces \cite{Herdeiro:2002ft}.
The reason was understood in \cite{Boyda:2002ba}: the uplifted spacetime
is a standard pp-wave, of the type that has been recently studied as the Penrose
limit of certain near horizon brane geometries \cite{Berenstein:2002jq}.
Such realization raised the hope that string theory could shed light on
one of the more important open problems in General Relativity, or more
generally, in gravitational theories: are geometries with closed timelike curves intrinsically
inconsistent or is propagation of matter
in these geometries intrinsically inconsistent?

A connection between these two lines of work was made in \cite{Drukker:2003sc}, where it was shown that
certain G\"odel-type spaces could be viewed as a background sourced by an over-rotating supertube
domain wall in the limit where the domain wall is taken to infinity. This work also gave support
to possible holographic interpretations of G\"odel spaces\cite{Boyda:2002ba, Brecher:2003rv, Nadav2}, since a
causal region of the  G\"odel-type space could be enclosed within a supertube domain wall that is not
over-rotating. Following \cite{Emparan:2001ux}, the authors of \cite{Drukker:2003sc} 
showed that the action for BPS supertube probes on the full G\"odel space contained kinetic terms that 
could become negative, and the 
same conclusions were drawn.

In the context of G\"odel-type spaces, 
recent results have called into question 
the conclusion that the appearance of these negative kinetic terms
implies an instability on the probe. In a U-dual language, it was found in \cite{BHH} that certain BPS probes
with negative kinetic terms were in fact stable because they sat at a maxima of an effective potential. This
agrees with the calculation of the quantum spectrum carried out in \cite{Russo:1994cv}, where there are no signs of ghost
or tachyonic string states.
In the supertube language, the classical stability of the probe can be inferred from the  
explicit solution \cite{B} to the probe equations of motion.

Given these results, it is natural to continue to look for pathologies in string theory on G\"odel-type spaces.
Another direction  of thought was considered in
\cite{B}, where closed geodesics were shown to exist in a certain G\"odel-type backgrounds. These
geodesics are not one dimensional lines, but rather the three dimensional worldvolumes of $D2$ branes.
Although their existence does not immediately imply that string theory cannot be well defined on these
backgrounds, there are general arguments which suggest that quantum theories cannot be consistently
defined under these circumstances, at least in the case of point particles.  
Because these backgrounds are highly supersymmetric, we
might expect that any potential pathologies arising from these closed geodesics 
will only appear in quantities that are not protected by
supersymmetry. Of course, these arguments would not apply to causal subspaces carved out by possible holographic screens, where
suitable boundary data could be given.
Recent discussions of related issues in string theory can be found in \cite{Boyda:2002ba, Harmark:2003ud, Dyson:2003zn, 1, Biswas:2003ku, Drukker:2003sc, Hikida:2003yd, Brecher:2003rv, BHH, 2, HT, 3, 4, 5, DI}.

The results of this paper are along the lines of \cite{B}. We argue that there exists closed supertube geodesics
on backgrounds sourced by over-rotating supertubes. Although these backgrounds can be ruled out physically, since
their matter content does not correspond to anything found in string theory, it is still useful to consider them in order
to uncover potential agents of chronology protection\cite{Hawking}. Our arguments are general 
enough to include certain backgrounds
sourced by uniformly smeared supertubes and some Kaluza-Klein compactifications\cite{Emparan:2001ux} 
transverse to the supertube,  
although they break down in the case of the domain wall background
constructed in \cite{Drukker:2003sc}. The same results are expected for backgrounds obtained through U-dualities.

The outline of the paper is as follows. In the next  section,
we provide a brief review of supertubes in flat space, followed by a review of supergravity backgrounds
sourced by supertubes. In Section 3, we consider supertube probes on these supergravity backgrounds. This section
is largely a repeat of the analysis carried out in \cite{B}, apart from some minor  complications. We prove in
Section 4 that the appearance of negative kinetic terms in the probe action does not lead to an instability of  
the BPS supertube probes, at least with respect to the zero frequency modes originally considered in \cite{Emparan:2001ux}. 
Furthermore, we show that the BPS energy is a upper bound on the Hamiltonian in the
regions where the kinetic terms are negative. 
In Section 5, we study the probe's dynamics, largely through an example. We show that closed geodesics exist in Section 6
using again a specific example, and then in the general case. In Section 7, we write down the probes
contribution to some of the components of the gravitational energy momentum tensor.  
Finally, in Section 8, we apply some of the techniques used in this paper to better understand
some features of the supertube geodesics on the G\"odel-type background discussed in \cite{B}. 
\newpage
\sect{Review}

In this section, we review some results from \cite{MT, Emparan:2001ux} about supertubes in flat space and
supergravity backgrounds sourced by supertubes.
\subsection{Supertubes in Flat Space}
It was shown in \cite{MT} that cylindrical $D2$ branes
can be supported against collapse by angular momentum generated by
electric and magnetic fields on their worldvolume, and that these
supertubes are $\frac{1}{4}$ BPS. 
To be specific, let us
consider a cylindrical $D2$ brane configuration extended in a
direction $y$ and wrapping some angular coordinate, $\theta$,
$N$ times at a radius $r$. If we restrict attention to the $U(1)$
zero mode components of the worldvolume fields, the system is
described by a Born-Infeld Lagrangian. 
\bequ 
{\cal{L}} =
-|N| \sqrt{(1-v^2 )( {r^2} + B^2 ) - E^2 {r^2}} \ , 
\eequ 
where $E=F_{0y}$ and $B=F_{y \, \theta }$ are
the non-zero components of the field strength and
\bequ
v^2 = \dot{\rho}^2 + \dot{r}^2 \ .
\eequ
Here,  $\dot{r}$ is the time derivative of the radius of the supertube, while $\dot{\rho}$ is the velocity
transverse to the $r, \theta$ plane. 
It is natural to remove the dependence on the electric field in favor of
the conserved conjugate momentum,
\bequ
\Pi = N^{-1} \frac{\partial {\cal L}}{\partial E} \ , 
\eequ
by considering the Routhian ${\cal R} = {\cal L} - (N\Pi) E$,
\bequ
{\cal R} = -\frac{|N|}{r}\sqrt{(1-v^2 )(r^2 + \Pi^2)(r^2 + B^2)} \ .
\label{flatR}
\eequ
A $D2$-brane system with nonzero
field strength can be thought of as a bound state of $D2$-branes,
$D0$-branes, and fundamental strings. The
momentum conjugate to the electric field,
$N\Pi$, is just the number of strings that are extended along $y$, and  $NB$ is
the number of $D0$-branes per unit length in the $y$ direction.
\bequ
q_{F1} = N\Pi \ , \ \ \ \ q_{D0} = NB
\eequ
The stationary configurations are supersymmetric and obey the BPS
conditions 
\bequ 
r^2_{BPS} = |\Pi B|  \ , \ \ \ \  {\cal{H}}_{BPS} =
|N\Pi| + |NB| \, . 
\eequ 
where ${\cal H} = -{\cal R}|_{v^2=0}$ is the Hamiltonian of
the stationary system. 

The supertube carries an angular momentum per unit
length given by
\bequ 
{J} = -N\Pi B = -\frac{ q_{F1} q_{D0}}{N} \ .
\eequ 
It was also shown in \cite{Emparan:2001ux}  that properly oriented strings, extended in the $y$ direction, and $D0$-branes
preserve the same supersymmetry as the supertube. In particular, the charges of the string and $D0$-brane must
be of the same sign as those of the supertube. Then we can consider a supertube of the type just described
with additional string and $D0$ charges which do not contribute to the angular momentum. We can also consider
the superposition of supertubes of the same stationary radius. These considerations lead to the   
following
bound on the angular momentum of any composite system.
\bequ 
J^2 \le R^2 |Q_{F1} Q_{D0}| \ ,
\label{bound} 
\eequ
where $Q_{F1}$ and $Q_{D0}$ represent the total charges of the system, and $R$ is the stationary radius.

\subsection{Supertube Sourced Backgrounds}
We now turn to the supergravity description of supertubes constructed in \cite{Emparan:2001ux} and
begin by recalling the family of one-quarter supersymmetric type IIA backgrounds considered
there.
\begin{eqnarray}
 \label{supertube}
    ds^2&=&-U^{-1}V^{-1/2}(dt-A)^2+U^{-1}V^{1/2}dy^2
      +V^{1/2}\delta_{ij}dx^idx^j \nn \\
    B_2&=&-U^{-1}(dt-A)\wedge dy+dt\wedge dy\,, \nn \\
    C_1&=&-V^{-1}(dt-A)+dt\,, \nn \\
    C_3&=&-U^{-1}dt\wedge dy\wedge A\,, \nn \\
    e^\phi&=& U^{-1/2}V^{3/4}\,.
\label{background}
\end{eqnarray}
Here $U$ and $V$ are harmonic functions in the eight dimensions
spanned by the $x^i$, and $A$ is a Maxwell field, which for supertube sourced backgrounds takes the form\footnote{
The notation is slightly different from that appearing in \cite{Emparan:2001ux}. In particular, $f'_{here} = -f_{there}$. 
This is
done so that we can
reserve the use of $f$ in order to make the notation in the coming sections  more closely match that found in \cite{B}.}
\bequ
A=-f'd\theta \ ,
\eequ
where $r$ and $\theta$ are polar coordinates on some plane spanned by two of the $x^i$, say $x_1$ and $x_2$.
The supergravity background of a single supertube is given by (\ref{supertube}) with\footnote{ We will assume $f'$ is 
positive, and therefore $J$ negative. The background then corresponds to a supertube with positive $N$, $\Pi$ and $B$, or
equivalently, positive $D2$, $D0$, and $F1$ charges where $dt\wedge dy$ and $dt \wedge dy \wedge d\theta$ define a positive
orientation. }
\begin{eqnarray}
f' &=& \frac{-J}{ \Omega} \, \frac{\left(r^2 + \rho^2 + R^2 \right) \, r^2}
{ 2\Sigma^5}  \,   ,
\nn \\
U &=& 1 + \frac{|Q_{F1}|}{ \Omega}  \,
\frac{\left(r^2 + \rho^2 + R^2 \right)^2 + 2 R^2 r^2}{6 \Sigma^5} \,, \nn \\
V &=& 1 + \frac{|Q_{D0}|}{ \Omega}  \, \frac{\left(r^2 + \rho^2 +
R^2 \right)^2 + 2 R^2 r^2}{6 \Sigma^5} \,, \label{choices}
\end{eqnarray}
where \bequ \Sigma (r,\rho) = \sqrt{\left(r^2 + \rho^2 + R^2
\right)^2 - 4 R^2 r^2}\,, \ \ \ \rho^2 = \sum_{i\ne 1,2} (x^i)^2
\label{defsigma} \eequ In the above expressions, $R$, $J$,
$Q_{D0}$, and $Q_{F1}$ are respectively the radius, angular
momentum, $D0$ charge, and $F1$ charge of the supertube sourcing
the background and $\Omega$ is the volume of the unit seven-sphere.
Multi-tube, smeared,  and domain wall solutions can be generated by
superposition \cite{Emparan:2001ux,Drukker:2003sc}. When the parameters of
this supergravity solution violate the bound (\ref{bound}) closed
timelike curves develop sufficiently close to the supertube. 
The relevant component of the metric takes the form
\bequ 
g_{\theta \theta} = V^{\frac{1}{2}}r^2 \, \left(1 - \frac{f'^2}{ 
UVr^2} \right)  \, , 
\label{l2} 
\eequ 
and the closed
curve generated by $\frac{\partial}{\partial \theta}$ becomes
timelike when 
\bequ
\Delta \equiv 1 -\frac{f'^2}{UVr^2} = 1 -f^2 r^2 < 0  \ .
\eequ 
Here we have also defined the positive quantity $f$, whose use will be  convenient in 
the following sections. By evaluating $\Delta$ at the location of source, 
\bequ
\Delta |_{r=R \, , \, \rho=0} = \frac{-J^2 + R^2|Q_{F1}Q_{D0}|}{  R^2|Q_{F1}Q_{D0}|} \ ,
\eequ
we see that closed timelike curves appear when the source is over-rotating. In fact, a more thorough
analysis \cite{Emparan:2001ux} shows that closed timelike curves appear only when the source is over-rotating.

Although the net $D2$ brane charge of the supertube source vanishes, there is a non-zero $D2$ dipole moment.
The number $N$ of $D2$ branes wrapping the $\theta$ direction can be calculated with the result
\bequ
N = -\frac{J}{R^2}
\eequ
which agrees with the probe result in the last section. 

When the supertube source does not violate the angular momentum bound (\ref{bound}), there is in principle a  microscopic
description of the system using $D$ branes as in the last section. On the other hand, when the bound is violated
we have no such description. There is more angular momentum in the system than can be accounted for by
crossed electric and magnetic fields on a supertube. We might consider the possibility that the extra angular momentum 
comes from excitations on the $D$ brane, but this would break further supersymmetries in the microscopic description. 
This suggests that
the over-rotating supertube supergravity backgrounds do not correspond to the supergravity background  of
any string theory matter.   
It also seems clear that any attempt to assemble a supertube system out of supertube matter 
which does not violate the angular momentum bound 
will result in a system that also does not violate the bound. Nevertheless, 
it is still useful to  ask if
string theory could possibly be well defined on these backgrounds, and in the process perhaps shed light on the larger question of 
chronology protection in string theory.

In the coming sections, we will be considering some specific backgrounds of the form (\ref{background}) as examples.
However, our arguments will not rely critically on the form of the functions (\ref{choices}). 
The facts that will be important to us, and that should be kept in mind are these: $(i)$ $\Delta$, $f$, $U/V$,
and $V/U$ are finite everywhere for nonzero background charges and $(ii)$ the functions $U$ and $V$ both diverge
only at the location of the source.
\section{Supertube probes}
We now turn to the study of supertube probes on the backgrounds
considered in the previous section. If we consider a cylindrical
$D2$-brane wrapping the $\theta$ direction $N$
times, the system is described by a $U(N)$ gauge theory. Restricting attention to the $U(1)$
zero mode components of the field strength, transverse scalars and the radial mode, the
probe is described by a Born-Infeld Lagrangian with couplings to
the background $RR$ fields. 
\bequ 
{\cal L} = -|N\, | \,
e^{-\phi}\sqrt{-{\mbox d}{\mbox e}{\mbox t}(G + {\cal{F}})} -N\,
e^{\cal F} \sum C^{(n)} \, , 
\label{gen} 
\eequ 
where ${\cal F} = F
- B_{2}$ is the invariant field strength. 
For a system aligned with the source and centered at $r=0$, we can plug in the background (\ref{background}) 
to find\footnote{ Here, as in the rest of the paper, we set
$(2\pi)^2 L \tau_{2}$ = 1, where $\tau_{2}$ is the $D2$ brane tension at $g=1$, and $L$ is the compactification radius of the
$y$ direction. In the case that $y$ is not compact then ${\cal L}$ should be considered a Lagrangian density.}
\begin{eqnarray}
{\cal{L}}&=& -\frac{|N|}{(UV)^{\frac{1}{2}}} \sqrt{(
\dot{t}^2\Delta^{-1}-v^2)( {r^2}\Delta + UV^{-1}\bar{B}^2 ) -
U^2\bar{\cal E}^2 {r^2}\Delta } \nn \\
&&+\frac{\dot{t}N}{V}\bar{B} -\dot{t}NB+ \frac{Nf'}{V}{\cal E} \ ,
\label{Lag}
\end{eqnarray}
where,\footnote{Although the notation may not make it obvious, $\bar{B}$ and ${\cal E}$ are the non-zero components
of the invariant field strength ${\cal F}$.}
\bequ 
\bar{B} = B - U^{-1}f'
\ , \ \ {\cal E} = E + \dot{t}(U^{-1} - 1) \  , \ \ \bar{\cal E} =
{\cal E}  - \frac{f'\bar{B}\dot{t}}{UVr^2\Delta} \ , \ \ 
v^2 = UV( \dot{r}^2 + \dot{\rho}^2) \ ,
\eequ 
and
the dot indicates differentiation with respect to $\lambda$, which
can be interpreted as the worldvolume time coordinate. As in the flat space example of Section 2,
$E$ and $B$ are the electric and magnetic fields on the brane and
we choose to consider only configurations\footnote{ Choosing
temporal gauge (on the probe) $A_0 = 0$, one must enforce the
Gauss law constraint $\partial_\theta \Pi_\theta + \partial_y
\Pi_y = 0$. It is consistent to set $F_{0 \, \theta} = \Pi_\theta
= 0$ for vanishing $\partial_y \Pi_y$. On the other hand, as
opposed to the flat space case, it is not consistent to fix $E =
F_{0y}$ to be a constant, since $E$ and $\Pi_y$ are not directly proportional
and differ by coordinate dependent terms.} with $F_{0 \,\theta} = 0$.
Also, we do not allow for center of mass motion in the $r,\theta$ plane. This is
consistent with the equations of motion since the background is invariant under
rotations in this plane. 

It will be useful to work with the Routhian, ${\cal R} = {\cal L} - (N\Pi) E$, which is given by
\begin{eqnarray}
{\cal R}&=& \frac{ - s^{\prime} |N|}{(UV)^{\frac{1}{2}} r^2  \Delta }\sqrt{ r^2  \Delta ( 
\dot{t}^2  \Delta^{-1}  -v^2)( r^2 \Delta+U^{-1}V\bar{\Pi}^2 )
(r^2 \Delta + UV^{-1}\bar{B}^2) } \nn \\
&&- \frac{\dot{t}Nf'}{UVr^2\Delta}(r^2\Delta + \bar{\Pi}\bar{B}) +\frac{\dot{t} N}{UV}(V{\Pi} + U{B}) -\dot{t}N(\Pi + B) \ ,
\end{eqnarray}
and which serves as a Lagrangian for $r$, $\rho$ and $t$. 
Here we have defined the momentum conjugate to $E$ as before, and have made use of the following further definitions.
\bequ
\Pi = N^{-1}\frac{\partial{\cal{L}}}{\partial E} \ , \ \
\bar{\Pi} = \Pi -V^{-1} f'  \ , \ \
s' = \mbox{sign}( r^2 \Delta+U^{-1}V\bar{\Pi}^2) \ .
\eequ
At this point, it is convenient to define some scaled quantities so that the Routhian
more closely resembles its flat space (\ref{flatR}) or G\"odel Universe \cite{B}  form. 
\bequ
{\cal N} = (UV)^{-\frac{1}{2}} N \ , \ \ {\cal P} = U^{-\frac{1}{2}} V^{\frac{1}{2}} \Pi\ , \ \ {\cal B} = U^{\frac{1}{2}} V^{-\frac{1}{2}} B 
\eequ
Then the Routhian becomes,
\begin{eqnarray}
{\cal R}&=& \frac{ - s^{\prime} |{\cal N}|}{ r  \Delta }\sqrt{(
\dot{t}^2 -v^2\Delta )( r^2 \Delta+ \bar{{\cal P}}^2 )
(r^2 \Delta + \bar{{\cal B}}^2) } \nn \\
&&- \frac{\dot{t}{\cal N}f}{\Delta}(r^2\Delta + \bar{{\cal P}}\bar{{\cal B}}) +{\dot{t} {\cal N}}({\cal P} + {\cal B}) -\dot{t}N(\Pi + B) \ , \label{Routhian}
\end{eqnarray}
where,
\bequ
\bar{{\cal P}} = {\cal P} -fr^2 \ , \ \ \ \bar{{\cal B}} = {\cal B} -fr^2 \ , \ \ \ \Delta = 1 -f^2 r^2 \ , 
\ \ \ s' = \mbox{sign}( r^2 \Delta+ \bar{\cal P}^2) \ .
\eequ
One can easily take the limit $f\rightarrow 0$ and $U,V\rightarrow 1$ to recover the flat space
Routhian (\ref{flatR}). By taking $f, U$ and $V$ constant, one recovers the the G\"odel Universe form found in \cite{B}. 

Now is a good time to point out an observation first made in \cite{Emparan:2001ux}; for a certain range of charges, in the
limit of small velocities the kinetic terms in the Routhian are negative. Setting $\dot{t}=1$ we find,
\bequ
{\cal R} = {\cal R}|_{v^2=0} + \frac{s'|{\cal N}|}{2r} \sqrt{( r^2 \Delta+ \bar{{\cal P}}^2 )
(r^2 \Delta + \bar{{\cal B}}^2) } \ v^2 + \dots
\label{expand}
\eequ
It would seem that whenever $s'=-1$ the kinetic terms are negative. However, the square root term in the Routhian (\ref{Routhian})
and in (\ref{expand})  is not even real unless $s'= s$, where 
\bequ
s = \mbox{sign}(r^2\Delta +  \bar{{\cal B}}^2) \ .
\eequ
Probes are  then forbidden\footnote{The field configuration of the probe in this region is roughly analogous to
probe in flat space with a super-critical electric field, where the Born-Infeld action becomes imaginary. In both
cases, with initial conditions that fix the 
square root to be real, the equations of motion do not allow it to become imaginary.} 
to enter or exist in regions where $s \ne s'$. In the regions where $s=s'=-1$, the kinetic 
terms are indeed negative, but as we will show in the next section this does not lead to any instabilities of BPS probes.

Since there is no explicit $t$ dependence the Routhian defines a conserved
energy $H = -\frac{ \partial {\cal R} }{ \partial \dot{t}}$ given by
\begin{eqnarray}
H &=& \frac{ s \, \dot{t}\, |{\cal N|} \sqrt{
    ( r^2 \Delta+\bar{{\cal P}}^2) 
    (r^2\Delta +  \bar{{\cal B}}^2)}}
{ r \Delta \left( 
\dot{t}^2  -v^2 \Delta  \right)^{\frac{1}{2}} } \nn \\  
%
&&+\frac{{\cal N}f}{\Delta}(r^2\Delta + \bar{\cal P}\bar{\cal B}) 
-{\cal N}({\cal P} + {\cal B}) + N(\Pi + B) \ .
\label{H}
\end{eqnarray}
Notice that the system is invariant under 
\bequ 
\dot{t}\rightarrow
-\dot{t} \, , \ \ \ N \rightarrow -N \, , \  \ \ H \rightarrow -H
\, . 
\eequ 
Physically, this means that a probe traveling forward
(backward) in time with energy $H$ can be interpreted as the charge
conjugate\footnote{ Flipping the sign of $N$ changes the sign of
the $D2$, $D0$, and $F1$ charges of the probe.} probe ($N
\rightarrow -N$) traveling backward (forward) in time with energy
$-H$. 
\sect{BPS Bounds and Stability}
In this section, we show that BPS supertube probes are stable despite the appearance of negative 
kinetic terms in the action. We further show that the BPS energy is an upper bound on the Hamiltonian
in the regions where the kinetic terms are negative.
Let us start with the Routhian (\ref{Routhian}) and define ${\cal H} = -{\cal R}|_{\dot{t} =1, v^2= 0}$, which is
the same as the Hamiltonian considered in \cite{Emparan:2001ux}, where for
positive $N$, $\Pi$, and $B$ the stationary solution was shown to
be supersymmetric, obeying the BPS conditions 
\bequ 
r^2_{BPS} = \Pi B \, \ \ \ {\cal H}_{BPS} = N\Pi + N B \ . 
\eequ 
We would first like to show
that the BPS conditions are satisfied whenever both $N \Pi $  and
$NB$ are positive. 
This should be expected on the following grounds. In flat space, the supersymmetries that are broken by a 
supertube of positive $N\Pi$ and $NB$ are the same as those broken by a $D0$ brane and
a fundamental string \cite{Emparan:2001ux}. The sign of the $D2$ charge does not make any difference in the matter. Thus a probe with
with the same $D0$ and $F1$ charges as a supertube source should break no further supersymmetries of the background,
independent of the relative sign of the $D2$ charge of source and probe.
We begin by writing ${\cal H}$ in a convenient form
\begin{eqnarray}
{\cal H} &=&  \frac{1}{\Delta r} \Bigl[ s \, |{\cal N}| \sqrt{{\cal P}{\cal B}} \Bigl(
    \left({\cal P} + {\cal B} -2f{\cal P}{\cal B} + ({\cal B}^{-1} -2f)(r^2-{\cal P}{\cal B}) \, \right) \nn
\\
&&    
  \ \ \ \ \ \ \ \ \ \ \ \ \ \ \ \ \ \ \,  \times \left({\cal P} + {\cal B} -2f{\cal P}{\cal B} + ({\cal P}^{-1} -2f)(r^2-{\cal P}{\cal B}) \, \right)\Bigr)^{1/2}  \nn
\\
&&\ \ \ \ \ \ \ - {\cal N} r \left( {\cal P} + {\cal B} -2f{\cal P}{\cal B} - 
f(r^2 - {\cal P}{\cal B}) \right)\Bigr] + N(\Pi + B) \ .
\label{rcalH}
\end{eqnarray}
When $r^2 = {\cal P}{\cal B} = \Pi B$, it is not hard to see that ${\cal H} = N(\Pi +B)$ using the fact that
\bequ
s\,|_{r^2 ={\cal P}{\cal B}} = s'|_{r^2 ={\cal P}{\cal B}} = 
\mbox{sign}\left(  {\cal N}({\cal P} + {\cal B} -2f{\cal P}{\cal B})\right)|_{r^2 ={\cal P}{\cal B}} \ ,
\eequ 
when both $N \Pi $  and
$NB$ are positive. We can also show that this point is a stable extrema by expanding ${\cal H}$ to second order.
This is not as tedious as it may seem, since we only need to expand the explicit $r$ dependence appearing
inside the square brackets of (\ref{rcalH}) to obtain
\bequ
{\cal H} = N(\Pi + B) + \left. \frac{2s\, |{\cal N}|}{|{\cal P} + {\cal B} -2f{\cal P}{\cal B}|}
\left(\frac{\sqrt{{\cal P}{\cal B}}}{r}\right) \left( \frac{1-{\cal P}{\cal B}f^2}{\Delta}\right) \right|_{r^2 ={\cal P}{\cal B}}(r-\sqrt{{\cal P}{\cal B}})^2 + \dots
\eequ
The two factors in large parentheses evaluate to one, and we
see that the extrema
of ${\cal H}$ is a maximum when $s=-1$, and a minima otherwise.  Since the kinetic term is also proportional
to $s$, all the BPS solutions are in fact stable. Since the above result is not
dependent on the precise form of $U$, $V$ and $f$ which define the background, the
result will hold in a broad class of backgrounds including certain G\"odel spaces,
where the same result \cite{BHH, B} was obtained. 

We now set out to show a stronger result; for probes with $N>0$ the BPS energy is an upper bound
on the Hamiltonian whenever $s=s'=-1$. The fact that $s$ and $s'$ are equal to minus one requires that both
${\cal P}$ and ${\cal B}$ are greater than zero, so we now restrict our attention to probes with these charges. 
Let us undo the massaging of (\ref{rcalH}),
\bequ
{\cal{H}} - N(\Pi + B) = X + Y \ ,
\label{Ham5}
\eequ
where
\begin{eqnarray}
X &=& \frac{ s \, |{\cal N|}}{r \Delta} \sqrt{
    ( r^2 \Delta+\bar{{\cal P}}^2) 
    (r^2\Delta +  \bar{{\cal B}}^2)} \ , \\
Y &=& \frac{{\cal N}f}{\Delta}(r^2\Delta + \bar{\cal P}\bar{\cal B}) 
-{\cal N}({\cal P} + {\cal B}) \ .
\end{eqnarray}
Then, it is possible to show
\bequ
X^2 - Y^2 = \frac{{\cal N}^2(r^2-{\cal PB})^2}{r^2 \Delta} \ .
\eequ
Whenever $\Delta>0$ we have $X+Y \ge 0$, since $X$ is positive and has
a magnitude greater than or equal to that of $Y$. So, in this region the BPS condition is a lower
bound on the energy. On the other hand, when $\Delta<0$, the sign of $Y$ determines whether
we have an upper or a lower bound on ${\cal H}$. Let us rewrite $Y$.
\begin{figure}
\begin{picture}(0,0)(0,0)
\put(-3,185){$x^i$}
\put(-3,102){$0$}
\put(98,-10){$R$}
\put(127,-10){$r_{BPS}$}
\put(210,-5){$r$}
\put(131,80){${\cal S}'$}
\put(112,129){${\cal S}$}
\put(203,122){${\cal S}_\Delta$}
\end{picture}   
\centering\epsfig{file=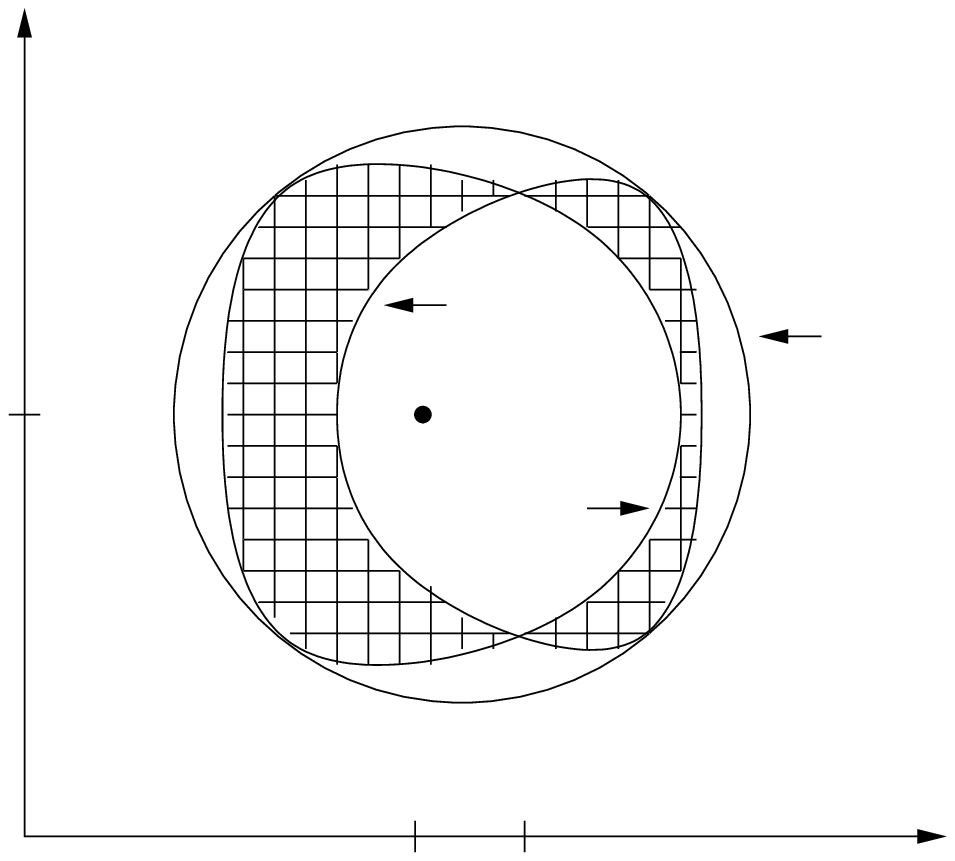,width=8cm}   
\caption{Sketch of the surfaces ${\cal S}_\Delta$, ${\cal S}$ and ${\cal S}'$ which are defined by the
zeros of $\Delta$,  $( r^2 \Delta+\bar{{\cal B }}^2)$, and $( r^2 \Delta+\bar{{\cal P }}^2)$ respectively,
for a case when the BPS radius intersects the $s=-1$ domain. The surface ${\cal S}_\Delta$ depends only on the
background, whereas ${\cal S}$ and ${\cal S}'$ also depend on the probe under consideration. Probes cannot exist
in the shaded region between  ${\cal S}$ and ${\cal S}'$. These three surfaces at a fixed time are topologically $S^6\times 
S^1 \times R^1$. What is sketched is the cross section for fixed $\theta$, $y$ and five of the six $x^i$. }
\end{figure}
\bequ
Y =  \frac{{\cal N}}{(-\Delta)}  \left( {\cal P} + {\cal B}  - 
f(r^2 + {\cal P}{\cal B})\right) \ . 
\eequ
Using this is expression along with the definitions of $s$ and $s'$, one can show that whenever $N>0$ and $s=s'=-1$, 
the sign of $Y$ is also negative, and therefore the
BPS energy is an upper bound on the Hamiltonian. The BPS bound is also saturated at the location of the
source where $U$ and $V$ diverge and ${\cal N}$ vanishes. 

For a given probe the background can be divided into domains
which are separated by impenetrable barriers, in the form of infinite potentials or surfaces where
 $(r^2 \Delta+\bar{{\cal P}}^2) = 0$ or $(r^2\Delta +  \bar{{\cal B}}^2) = 0$. It is natural to conjecture
that on each of these domains the BPS energy is a bound on the Hamiltonian, but we have not found a simple
proof of this. Although this point is not critical for our purposes, 
let us see how it works by considering the example sketched in  Figure 1, for a probe with positive
$N$, $\Pi$, and $B$. Near the surface ${\cal S}_\Delta$, where $\Delta=0$, the Hamiltonian (\ref{Ham5}) is
potentially divergent. The relevant terms for nonzero $\bar{\cal P}$ and $\bar{\cal B}$ can be can be written 
\bequ
{\cal H} \sim \frac{|{\cal J}| - fr{\cal J} }{r(1+fr)(1-fr)} \ .
\label{diverge}
\eequ
Here, ${\cal J}= -{\cal N} \bar{\cal P} \bar{\cal B}$ is proportional to the angular momentum, and so 
we see that ${\cal H}$ diverges at ${\cal S}_\Delta$ if the probes angular momentum is negative. 
If ${\cal J}$ is positive then ${\cal H}$ is finite, and since ${\cal H}$ is finite at $r_{BPS}$, the
angular momentum of the probe must be positive at the intersection of $r=r_{BPS}$ and ${\cal S}_\Delta$. If we think
about moving the probe on the surface ${\cal S}_\Delta$, the only way the angular momentum can change sign
is if $\bar{\cal P}$ or $ \bar{\cal B}$ becomes zero first. At these points the surface  ${\cal S}$ or ${\cal S}'$
must intersect ${\cal S}_\Delta$ as shown in Figure 1. 
Then it is possible to define domains which will be bounded by the union of 
certain subsets of ${\cal S}$ and  ${\cal S}'$ along with subsets of ${\cal S}_\Delta$ on which ${\cal H}$ diverges.
As we will see in the next section, although the Hamiltonian is 
finite on ${\cal S}$ and ${\cal S}'$, the dynamics of the probe do not allow it to cross these surfaces. 
\sect{Probe Dynamics}
Given the form of the action, it does not seem practical to solve
for the probe trajectories.
Instead,  we will try to simplify the problem as much as possible
in an attempt to understand some general features of the probes motion. 
We will consider only radial oscillations 
by requiring that
$\rho = \dot{\rho} = 0$, which is consistent with the equations of motion since the non-derivative
part of the action depends only on $\rho^2$.
Next, we will find it useful to define some potentials through the expression
\bequ
H = \frac{\dot{t}}{(\dot{t}^2 - UV\dot{r}^2 \Delta)^{\frac{1}{2}}} \ \Phi_{BI}(r)  + \Phi_t(r) \ ,
\label{H2}
\eequ
where the exact form of $\Phi_{BI}(r)$ and $\Phi_t(r)$ can be read off from (\ref{H}). 
Much of the dynamics can be understood just from finding the various turning points. The radial turning points
occur when
\bequ
H = \Phi_{\pm}(r) \equiv  \pm \Phi_{BI}(r) + \Phi_t(r) \ ,
\eequ
where the sign is determined by the direction of time flow. 
It is useful to keep in mind that  $\Phi_{+}(r)$ is identical to the Hamiltonian considered in the 
last section (\ref{Ham5}) once $\rho$ is set to zero. That is, $\Phi_{+}(r) = {\cal H}|_{\rho = 0}$. 
The temporal turning points, where $\dot{t}=0$, occur at radii determined
by
\bequ
H = \Phi_t(r) \ .
\eequ
To better understand this, let us consider 
Figure 2 which shows a sketch  $\Phi_{\pm}$ and $\Phi_t$ for a over-rotating background with $U=V$ and probe charges 
$N, {\cal B} = {\cal P} >0$. All the features of this sketch can be deduced based on the 
explicit form of the potentials. The potentials $\Phi_\pm$ blow up at $r=0,\infty$. When $\Delta=0$, marked
by the vertical lines in the sketch, $\Phi_-$ is finite since it corresponds to minus the Hamiltonian of
a probe with positive angular momentum, whereas $\Phi_+$ diverges since it corresponds to the Hamiltonian of
a probe with negative angular momentum. The BPS maximum occurs at $r^2 = \Pi B$, where $\Phi_+ = N(\Pi + B)$.
The over-rotating source is located at $r=R$, and at this point all three potentials are equal to
$N(\Pi + B)$ since ${\cal N}$ vanishes there. The only other places that all the
potentials are equal occur where $s=s'$ changes sign, which occurs twice.  The fact that there are no other
important features to the sketch is a numerical result for some particular choices of charges.

\begin{figure}
\begin{picture}(0,0)(0,0)
\put(12,190){$\Phi$}
\put(75,190){$(a)$}
\put(194,110){$(b)$}
\put(342,35){$(c)$}
\put(-4,151){\(\scriptstyle{{\cal H}_{BPS}}\)}
\put(-4,141){\(\scriptstyle{H_{min}}\)}
\put(3,92){\(\scriptstyle{H_{0}}\)}
\put(180,10){$r_{BPS}$}
\put(158,10){$r_{0}$}
\put(247,8){$R$}
\put(272,6){$\Phi_+$}
\put(368,6){$\Phi_-$}
\put(62,6){$\Phi_-$}
\put(445,10){$r$}
\end{picture}  
\centering\epsfig{file=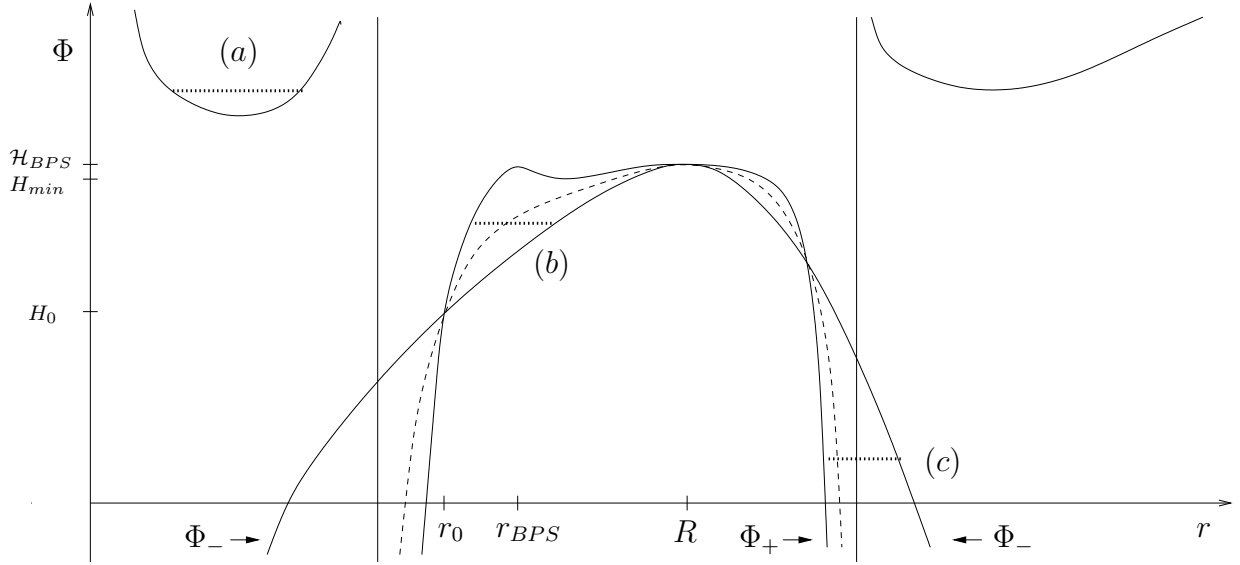,width=16cm}   
\caption{Sketch of effective potentials $\Phi_\pm$ (solid lines) and $\Phi_t$ (dashed line) for a background with $U=V$, and
a probe with charges $N, \Pi=B>0$, whose BPS radius lies within the $s=-1$ region. The vertical lines occur where $\Delta=0$. 
The point $r_0$, where all the potentials are equal, marks one of the radial positions of surface ${\cal S}={\cal S}'$.
The other position occurs to the left of the source, where all the potential are also equal.
The horizontal (dotted) lines labeled $(a)$, $(b)$, and $(c)$ represent possible configurations
of radial oscillation. }
\end{figure}
For a given $H$, we can simply draw the horizontal line level with $H$ and find two
radial turning points. 
Trajectory $(a)$ is a situation with which we are familiar. 
The radius oscillates between two turning points in the potential well\footnote{This trajectory is actually unstable
in the sense that the minimum of the well is a saddle point if the $\rho$ dependence is taken into account}
of $\Phi_+$. In the regions where $\Delta<0$, the radius can 
oscillate between a solution of $H=\Phi_+$ and $H=\Phi_-$, changing its direction of time
flow somewhere between when $H=\Phi_t$. Trajectory $(b)$ describes such a scenario. We can think of this trajectory
as the natural evolution after placing a future directed probe with zero radial velocity in the $s=-1$ region. 
Since the radial kinetic term is negative, the radius initially increases 
as it climbs the effective potential\footnote{$\Phi_+$ and $\Phi_-$ define effective potentials for future and past directed probes respectively. These potentials are only useful guides for understanding the motion when $\dot{r}$ is small, and the square root of the Routhian (\ref{Routhian}) can be expanded into what look like non-relativistic kinetic and potential terms.} 
$\Phi_+$.
At some point it crosses $\Phi_t$ 
and changes its direction of time flow as the radius continues to increase.
Next, it reaches the radial turning point determined by $H=\Phi_-$, 
after which the radius begins decreasing. 
Lastly, a probe following trajectory $(c)$ oscillates into and out of the $\Delta<0$ region.
However, when crossing the surface ${\cal S}_\Delta$, the probe with positive $N$ is always traveling
backward in time. This is consistent with the fact that probes of negative angular momentum cannot cross ${\cal S}_\Delta$.
Notice that there are no trajectories which allow the probe to
cross the surface ${\cal S}'$ or ${\cal S}$, which are the same in Figure 2. Even when $\rho\ne 0$, these
surfaces represent impenetrable barriers for the probe. 

We conclude this section by noting that if $\rho$ is not set to zero one must solve
the second order differential equations obtained from the Routhian. One the other 
hand, once $\rho$ is set to zero, the radial motion can be solved simply by finding
two radial turning points $r_1$ and $r_2$, fixing the reparametrization invariance of the action by setting
\bequ
r =  \frac{(r_1 + r_2)}{2} - \frac{(r_2 -r_1)}{2} \, \mbox{sin}\lambda \ ,
\eequ
and then integrating
\bequ
\dot{t} =  \frac{s\,  (UV)^{\frac{1}{2}} \Delta (H-\Phi_t) \, |\dot{r} |}{(\Delta(H-\Phi_t)^2 - \Delta \Phi_{BI}^2)^{\frac{1}{2}}} \ ,
\label{tdot2}
\eequ
which is the result of solving (\ref{H2}) for $\dot{t}$. 
Given this parameterization, the radial oscillations have period $2\pi$. It is convenient to define a time drift $T$ through one cycle of motion,
\bequ
T = \int_0^{2\pi} d \lambda \, \dot{t} \ , 
\label{T}
\eequ
which is of course independent of parameterization. In the case that $T=0$, the coordinate $t$ is also periodic and
the geodesic closes.
\sect{Closed Geodesics}

We would now like to argue for the existence of closed geodesics. 
We will do so by finding geodesics pairs, which we define as a pair of geodesics
that are continuously connected to one another, via the energy $H$, such that one has $T>0$ while the other has $T<0$.
Since for continuously connected geodesics, $T$ is also continuous, we will conclude 
there must exist a closed geodesic with $T=0$. We begin by again considering a background with $U=V$ and a probe with
charges $N, \Pi = B>0$ whose BPS radius lies within the $s=-1$ region as in Figure 2. In this sketch, geodesics are
represented as horizontal (dotted) line segments with endpoints on the graph of $\Phi_+$ or $\Phi_+$. 
As we vary the energy $H$ of a geodesic  
we can accordingly raise or lower these line segments on the graph. Smoothly connected line segments, that do not
join with  or split\footnote{As the representation of the geodesic approaches an unstable extrema on a graph, 
it either splits into two 
disconnected segments, or the reverse process occurs, where two segments join. Geodesics connected by the energy $H$ are not
continuously connected if such a process occurs at an intermediate value of $H$.}  
into other line segments,
will then represent continuously connected geodesics. In what follows, the word \lq geodesic' or \lq trajectory' may,
strictly speaking, only
refer to this representation as a line segment. 
After discussing the example sketched in Figure 2, we will move on to the general case. 
\subsection{An Example}
We will first find a geodesic with $T<0$ by focusing on trajectories similar to $(a)$, shown in Figure 2, 
in the limit we approach the surface ${\cal S}$.
Let us denote the radial position of ${\cal S}$ as $r_0$  and the value of the potentials at that point
$H_0$. 
Then we can approximate the system by only taking the linear terms
in the potentials
\bequ
\Phi_{BI} = b (r - r_0) + {\cal O}(r-r_0)^2 \ , \ \   \Phi_t= H_0 + a (r-r_0) + {\cal O}(r-r_0)^2 \ .
\eequ
Given the restriction on the charges of the background and probe, it is possible to show that $a>b>0$.
For $H$ near $H_0$, the radial turning points occur at
\bequ
r-r_0 = (H-H_0) \frac{ a \pm b}{a^2 - b^2}  + {\cal O}(H-H_0)^2 \ .
\eequ
We can fix the reparametrization invariance by setting
\bequ
r-r_0=(H-H_0) \frac{ a - b \, \mbox{sin} \lambda}{a^2 - b^2} \ .
\eequ
Then we find that
\bequ
\frac{|\dot{r} |}{(-(H-\Phi_t)^2 + \Phi_{BI}^2)^{\frac{1}{2}}} = C^2 \ ,
\eequ
where $C^2$ is a positive constant. Then using (\ref{tdot2}) and (\ref{T}) we find 
\bequ
T = (H-H_0)(-s) C^2 \sqrt{-UV\Delta}\, |_{r_0}\int_0^{2 \pi} d\lambda 
\frac{( -b^2 + ab \, \mbox{sin}\lambda)}{a^2-b^2} + {\cal O}(H-H_0)^2 \ .
\eequ
Since all the factors outside the integral are positive,
the drift in time $T$ through one cycle of motion is negative in the limit $H \rightarrow H_0$ from above.

In order to find a geodesic with $T>0$, we consider again trajectories similar to $(a)$, but this time we focus
on trajectories with $H$ near but smaller that $H_{min}$, where $H_{min}$ is defined as the value of the minimum
of $\Phi_+$ located between $r_{BPS}$ and $R$ in Figure 2. 
To prove that the drift through time is positive it is convenient to split the trajectory into two segments, one
where $\dot{t}>0$ (left), and the other where $\dot{t}<0$ (right). 
Then on the first segment we can choose the parameterization
$t=\lambda$, and on the second $t=-\lambda$. As $H$ approaches $H_{min}$ it takes a finite  (negative) time for the
radius to complete the second segment of its trajectory. On the other hand, since the radius must \lq climb' and 
\lq descend'
the minimum\footnote{Apart from the sign changes of the kinetic and potential terms, the problem is analogous to
a ball rolling over a hill, where if the energy is adjusted sufficiently close to the balls stationary 
energy at the maximum of
the hill, the process can take an arbitrarily long time. To see this, we can turn figure 2 upside down,
forget about the negative sign of the kinetic term, and think of the ball moving over the hill 
of (minus) $\Phi_+$.} 
during its first segment, by adjusting $H$ sufficiently close to $H_{min}$ we can make the time
interval of the first segment arbitrarily large.  Thus we have found a geodesic with $T>0$, which is 
continuously connected to another geodesic with $T<0$. This geodesic pair ensures the existence of a
closed geodesic. 

We will use these arguments again so let us summarize. When attempting to
identify geodesic pairs by sight, it is natural to look near the 
stable or unstable extrema of the potentials $\Phi_\pm$, where
the sign of $T$ can easily be determined to be the same sign as the subscript of $\Phi$. 
For these purposes, the potential configuration at $r_0$ in Figure 2 can
also be considered an extrema, where as we have shown, $T$ is negative. 
\subsection{General Backgrounds}
For the general over-rotating  background, we will now argue that it is always possible to find closed geodesics. 
The first step is to find a probe whose BPS radius lies in the $s=-1$ region as in Figure 1, which requires
$N$, $\Pi$, and $B$ all be positive. 
Since the background is over-rotating, $\Delta$ is negative in some region. This allows us to
choose a radius $r_{BPS}$, not equal to $R$, such that
\bequ
\Delta|_{r=r_{BPS}\, , \, {\rho=0}} < 0 \ .
\label{del3}
\eequ
Next we need to choose the probe charges, $\Pi$ and $B$, so that $\Pi B = r_{BPS}^2$ and
\bequ
s\,|_{r^2 ={\cal P}{\cal B}\, , \, {\rho=0}} = s'|_{r^2 ={\cal P}{\cal B}\, , \, {\rho=0}} = 
\mbox{sign}\left(  {\cal N}({\cal P} + {\cal B} -2f{\cal P}{\cal B})\right)|_{r^2 ={\cal P}{\cal B}\, , \, {\rho=0}} = -1 \ .
\label{sign3}
\eequ 
There exists a range of values that would satisfy these conditions, but for definiteness we can fix
$\Pi$ and $B$ by setting
\bequ
{\cal P}|_{r=r_{BPS}\,,\, \rho=0} = {\cal B}|_{r=r_{BPS}\,,\, \rho=0}  = r_{BPS} \ ,
\eequ
which satisfies (\ref{sign3}) since $fr_{BPS} >1$ as required by (\ref{del3}). 

The next step is to argue that what we know about the potentials for this probe 
is enough to deduce the existence of closed
geodesics. In the region between the source, located at $r=R$, and the surface ${\cal S}$ (or ${\cal S}'$)  
what we know is summarized in Figure 3a.
First, the potential $\Phi_+$ is bounded from above by ${\cal H}_{BPS}$ and the bound is saturated
only at $r_{BPS}$ and at $R$. Second, 
\bequ
\Phi_+ \ge \Phi_t \ge \Phi_- \ ,
\eequ 
with equality only at the source, $r=R$, and
on the surface ${\cal S}$  or ${\cal S}'$, whichever may be the case.
In fact, $\Phi_t$ is the average of the other potentials, but this point is not particularly important.
And lastly, the potentials are bounded from below as they are continuous functions in this region. 
\begin{figure}
\begin{picture}(0,0)(0,0)
\put(80,1){a)}
\put(225,1){b)}
\put(380,1){c)}
\put(35,22){${\cal S}$}
\put(80,23){$r_{BPS}$}
\put(138,22){$R$}
\put(-4,105){\(\scriptstyle{{\cal H}_{BPS}}\)}
\put(157,73){\(\scriptstyle{{H}_{e}}\)}
\put(2,120){$\Phi$}
\put(450,25){$r$}
\end{picture}  
\centering\epsfig{file=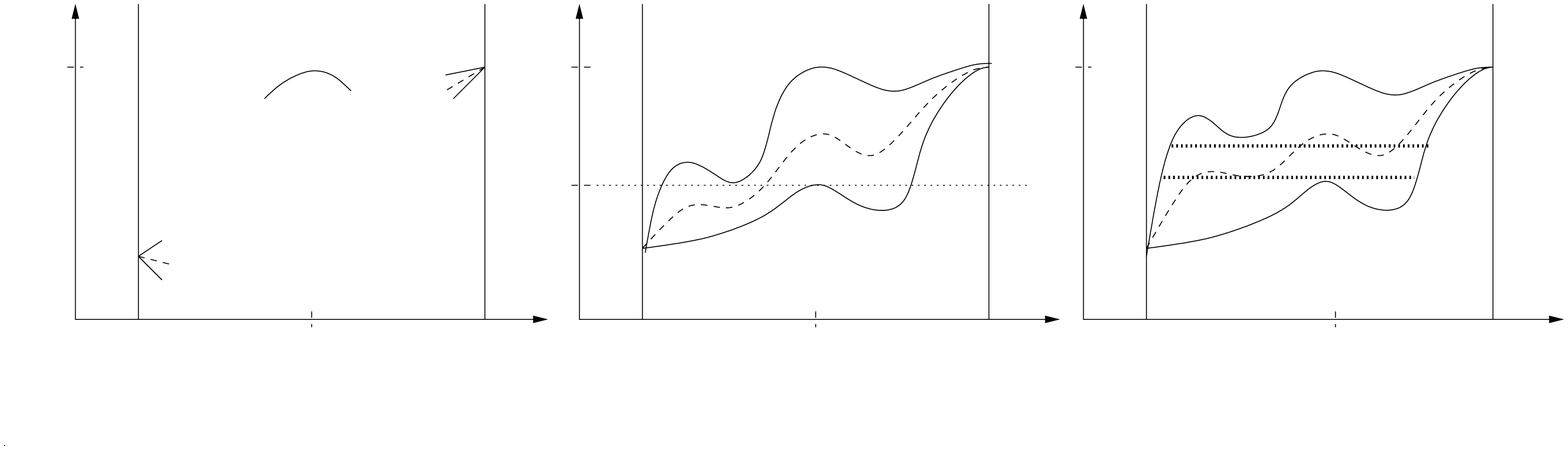,width=16cm}   
\caption{Sketches of possible forms of the potentials $\Phi_\pm$ (solid lines) and $\Phi_t$ (dashed line)  
for a probe whose BPS radius lies within the $s=-1$ region. We have assumed $r_{BPS}<R$, but if this is 
not the case the plots can be reflected through a vertical. The horizontal (dotted) lines in Figure c) represent
geodesics which together form a geodesic pair.}
\end{figure}

One would like to make the same argument as before. That is, find
a geodesic pair and conclude the existence of a closed geodesic.
Without knowing the explicit form of the potentials this would seem difficult, but in fact any generic completion of
the potentials in Figure 3a, subject to the constraints just outlined,  will contain 
geodesic pairs that can be identified by sight. Let us first discuss a counter example shown in Figure 3b. In some ways
this sketch is similar to the analogous region in Figure 2. We can identify two geodesics with 
opposite signs of $T$, one near the surface ${\cal S}$
and the other near the minimum of $\Phi_+$ located between $r_{BPS}$ and $R$. However,
this no longer constitutes a geodesic pair since the existence of the new extrema in $\Phi_\pm$ means that these two
geodesics are no longer continuously connected. Normally this would not be a problem, but when the values of the new extrema
are equal, as shown in Figure 3b, $T$ cannot be determined as $H$ approaches the new extrema, $H\rightarrow H_e$. 
On the other hand, if the values of these
extrema are shifted away from equality then geodesic pairs can again be identified, as shown in Figure 3c.  
We will not try to prove that the situation
sketched in figure 3b can never happen, rather we will argue that it is just not generic. 
We know that there is nothing that requires this coincidence because we have examples where
it does not occur; consider the case in Figure 2. So assuming that such a situation did occur
by considering another probe
with  a slightly different BPS radius and with slightly different charges we do not expect it to persist. 
Figure 4 shows a few more examples of generic potential configurations, and some geodesic pairs that can be identified.

Given a generic completion of the potentials in Figure 3a it is always possible to find a geodesic pair. A prescription
follows. Find the extrema of $\Phi_+$ with the lowest value. The geodesic with $H$ just below this value
will have $T>0$. Follow this geodesic
as $H$ is decreased.  
It must come upon an \lq extrema' of $\Phi_-$. 
This \lq extrema' may correspond to
an actual extrema, or to the potential configuration at $r_0$ in Figure 2. The geodesic with $H$ just above this
extremal value of $\Phi_-$ will have $T<0$. By construction these two geodesics
are continuously connected.

\begin{figure}
\begin{picture}(0,0)(0,0)
\put(80,1){a)}
\put(225,1){b)}
\put(380,1){c)}
\put(35,22){${\cal S}$}
\put(80,23){$r_{BPS}$}
\put(138,22){$R$}
\put(-4,105){\(\scriptstyle{{\cal H}_{BPS}}\)}
\put(2,120){$\Phi$}
\put(450,25){$r$}
\end{picture}  
\centering\epsfig{file=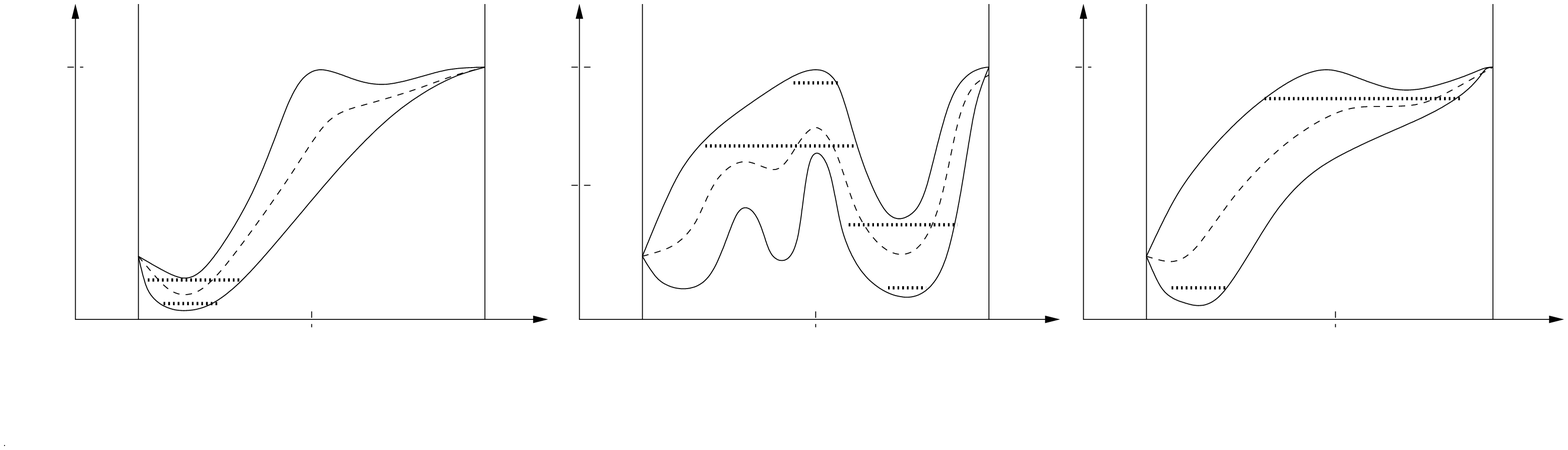,width=16cm}   
\caption{Sketches of possible forms of the potentials $\Phi_\pm$ (solid lines) and $\Phi_t$ (dashed line)  
for a probe whose BPS radius lies within the $s=-1$ region. The horizontal (dotted)
lines represent individual geodesics which go into making geodesic pairs.}
\end{figure}
We note that the only change observed\footnote{ Observation in this case means non-random biased
numerical sampling with low statistics using Mathematica.} 
in the form of the potentials relative to the case
in figure 2, is the possible existence of one extra minimum in $\Phi_-$ as sketched in Figure 4c.  
Before concluding this section, we must admit that we have cheated on one point. We assumed that the
source must lie in the $s=s'=-1$ region. 
If this were not the
case, we would have the BPS maxima plotted in Figures 3 and 4 surrounded by two surfaces of ${\cal S}$ or ${\cal S}'$,
rather than just one and the source. However, this would not affect the prescription just outlined in the previous
paragraph. Closed geodesics would generically exist in this case as well. 
The essential ingredients in this argument, which remain unchanged, are the existence of an extrema of $\Phi_+$,
and the fact that the all three potentials are equal at two points surrounding that extrema. When one of
those points is the source, the potentials are equal because $U$ and $V$ diverge there. When one considers
more general backgrounds, this may not be the case. In particular, for the supertube domain wall constructed in
\cite{Drukker:2003sc}, $U$ and $V$ approach a constant at the source, in which case these arguments break down.
\sect{Gravitational Couplings}

The probe's contribution to the energy momentum tensor can be calculated.
\bequ
T^{\mu \nu} = \frac{-2}{\sqrt{g}}  \frac{ \delta S }{ \delta g_{\mu \nu}} \ .
\eequ
It is convenient to define an energy momentum density ${\cal T}$ on the probe through the expression
\bequ
\sqrt{g} \ T^{\mu}_{\, \ \nu} = \int d^2\xi \, d\lambda \ {\cal T}^{\mu}_{\, \ \nu} \ \delta^{10}(X - X(\xi,\lambda)) \ ,
\eequ
where $X$ represents all the spacetime coordinates and $\lambda$ and $\xi$ parameterize the spacetime embedding of the
probe.
Then using the Lagrangian (\ref{gen}) one can show
\begin{eqnarray}
{\cal T}^0_{\, \ 0}  
&=& \dot{t}\left(H - {{\cal N}fr^2}  + {\cal N}({\cal P} + {\cal B}) - N(\Pi + B)\right)\ , \\
{\cal T}^0_{\, \ \theta} &=& \dot{t}(-N \bar{ \Pi }\bar{B}) \ .
\end{eqnarray}
In these expressions we have already fixed the $\xi$ parameterization, as we had done in the
Lagrangian (\ref{Lag}). 

When gravitational backreaction is ignored,
the supertube probe traveling along a closed geodesic leads to
divergent contributions to the energy momentum tensor. Of course,
this then invalidates the probe approximation, and indicates that any consistent treatment must take 
the effects of backreaction into account.

\section{G\"odel-type Universe}

\begin{figure}
\begin{picture}(0,0)(0,0)
\put(380,-5){$r$}
\put(145,-8){$r_{BPS}$}
\put(113,-8){$f^{-1}$}
\put(70,175){\(\scriptstyle{T>0}\)}
\put(70,120){\(\scriptstyle{T=0}\)}
\put(-22,23){\(\scriptstyle{{\cal H}_{BPS}}\)}
\put(-19,62){\(\scriptstyle{{H}_{\infty}}\)}
\put(-19,113){\(\scriptstyle{{2H}_{\infty}}\)}
\put(-10,170){$\Phi$}
\end{picture}  
\centering\epsfig{file=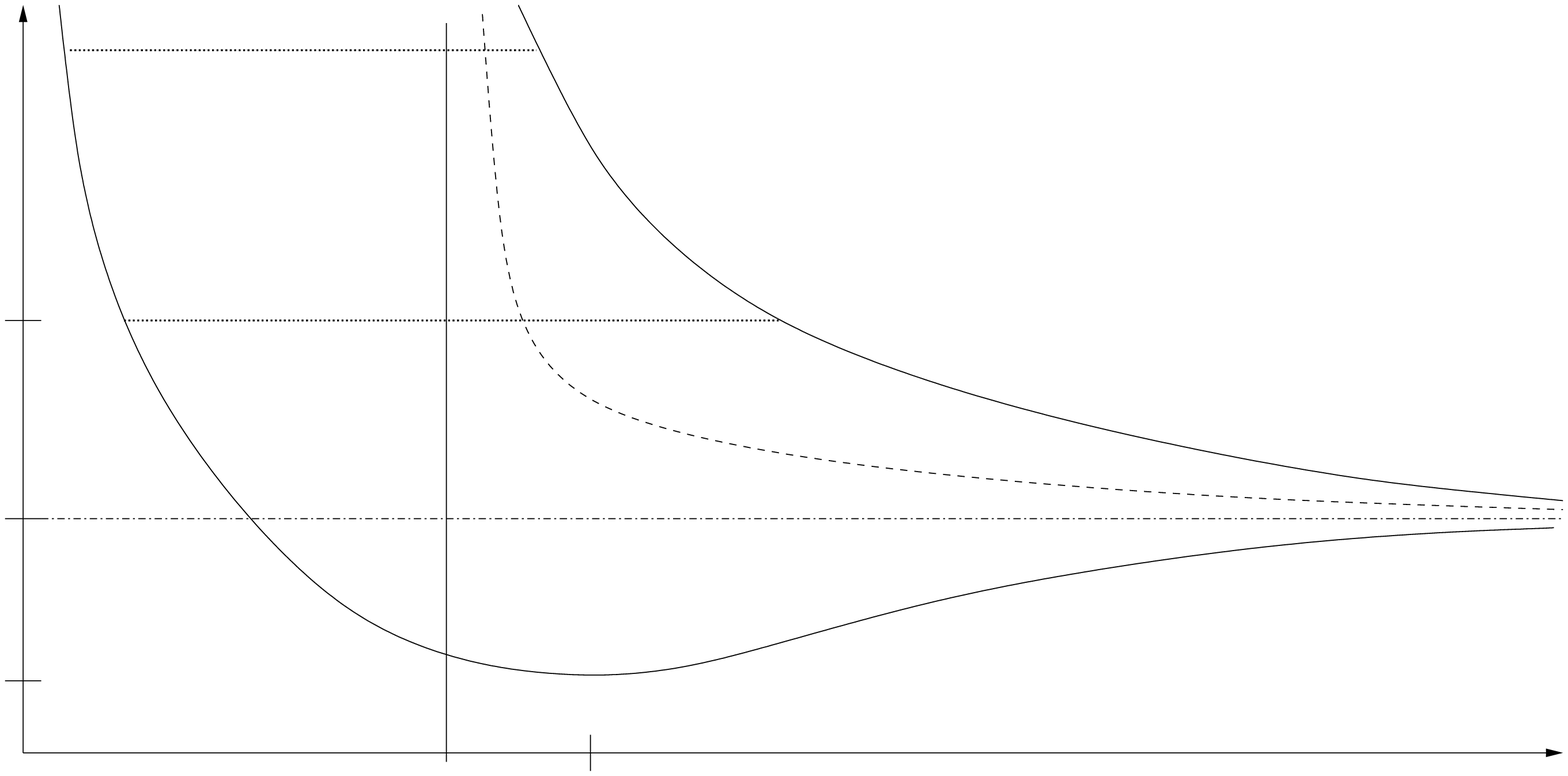,width=14cm}   
\caption{Sketches of potentials $\Phi_\pm$ (solid lines) and $\Phi_t$ (dashed line)  
in a G\"odel-type background, for a probe with $N$, $\Pi=B \, < 0$.}
\end{figure}

The techniques used in the previous sections can also be used to study the  G\"odel-type background considered in 
\cite{Drukker:2003sc,B}. However,
using these techniques we cannot conclude that closed geodesics exist. Nonetheless, many features of the 
explicit solution\cite{B} of the probes motion can be more easily understood in terms of a sketch of the potentials. 
As shown in \cite{Drukker:2003sc}, to obtain a G\"odel-type space from the background 
(\ref{background}) we just need to set $U=V=1$ and
$f$ to be a constant. Figure 5 shows a sketch of the potentials for a probe with $N$, $\Pi=B \, <0$. In \cite{B}, it
was shown that closed geodesics exist for probes with these charges. Note that these charges have opposite
signs relative to the ones we have been focusing on in this paper. In particular, when $\Pi$ and $B$ are less than
zero, $s$ and $s'$ are always one. Thus there is no region where the kinetic terms become negative. 

As shown in the figure, there is a stationary solution which is BPS, and the potentials at large $r$ all approach
\bequ
H_\infty \equiv - Nf^{-1} + N(\Pi + B),
\eequ
with $\Phi_t$ and $\Phi_-$ approaching from above and
$\Phi_+$ from below. 
Whether or not the potentials diverge at the velocity of light surface, where $\Delta=0$,  can be determined
using the expression (\ref{diverge}). In the case at hand, $\Phi_+$ is finite, while $\Phi_-$ diverges.
For energies ${\cal H}_{BPS}<H<H_\infty$, the radius oscillates in the potential well of $\Phi_+$ and the probe
never changes its direction of time flow. When $H=H_\infty$, the radial oscillation becomes infinitely large and
the radial motion is no longer periodic. We can think of this situation as a probe of large radius in the distant past which
contracts until it reaches its radial turning point.
At this point, it begins to expand and does so for the rest of its future.
For geodesics with energies $H_\infty<H<2H_\infty$, the solution \cite{B} to the equations of motion shows that
$T$ is always less than zero. In fact, $T$ diverges negatively as we approach $H_\infty$ from above. This is more
or less understandable from the sketch, since  such a probe would spend most of its proper time out at large radii
where $\dot{t}$ is negative. At very large energies, one might guess that $T$ is again positive based on the sketch, but
one must solve the equations of motions to prove that this is true. Finally, when $H=2H_\infty$, $T$ vanishes. But again,
to prove this requires more than the sketches. 

Since the effective potential $\Phi_+$ is finite at infinity, we might naively guess that probes with energy larger
than this asymptotic value would escape to infinity. However, from Figure 5 we see that probes with large energy
will in fact change their direction of time flow. Once this happens, it is perhaps more natural to 
think of this as a charge conjugate probe traveling forward in time. In this case, one should use the effective
potential $-\Phi_-$ to determine the radial turning points. In this way, we see that the typical motion of the
probe is bounded and periodic in $r$.

\section *{Acknowledgments}
We would like to thank Vika Naipak for her much needed assistance with Mathematica. We are especially grateful to Oren Bergman
and Shinji Hirano for useful discussions.
This work was supported by the Israel Science Foundation under grant No. 101/01-1.

\end{document}